\documentclass[preprint,prl,twocolumn,10pt]{revtex4}%

\usepackage{amssymb}
\usepackage{amsmath}
\usepackage{graphicx}

\begin{document}

\title{On the cosmological gravitational waves and cosmological distances}
\author{V.A. Belinski}
\affiliation{ICRANet, Piazza della Repubblica, 10, 65100 Pescara, Italy}
\affiliation{IHES, 35, Route de Chartres, F-91440 Bures-sur-Yvette, France}
\author{G.V. Vereshchagin}
\affiliation{ICRANet, Piazza della Repubblica, 10, 65100 Pescara, Italy}
\affiliation{ICRANet-Minsk, National Academy of Sciences of Belarus, \\ Nezavisimosti av. 68, 220072 Minsk, Belarus}


\begin{abstract}
We show that solitonic cosmological gravitational waves propagated through the Friedmann universe and generated by the inhomogeneities of the gravitational field near the Big Bang can be responsible for increase of cosmological distances.
\end{abstract}

\maketitle

\section{Introduction}

At present the most popular cosmological model --- $\Lambda$CDM model successfully explains a wealth of cosmological observations. However, it involves a hypothetical substance, ``dark energy'', having unusual physical properties. According to interpretation of various data, there is a need for accelerated expansion in the recent history of the universe, and hence dark energy is required to dominate the energy budget of the universe.

So far the main cause of introduction of dark energy was the discrepancy between observations of distant type Ia supernovae \cite{P, R} and Friedmann cosmological models with ordinary matter. Additional arguments include: tension between age estimates of globular clusters \cite{Krauss} and the age of the Universe, determined thanks to the measurement of the Hubble parameter \cite{Riess16,Planck,LIGO} measurement of the baryon acoustic oscillations signature \cite{BAO} providing the ratio of absolute distances in different cosmological epochs, and the inference from X-ray observation of massive galaxy clusters of the evolution of their number density \cite{Xray}, all consistent with recent accelerated expansion. Moreover, measurements of anisotropy spectrum of the cosmic microwave background \cite{Boom,Planck} suggest that observed universe is nearly spatially flat, while accounting for matter usually gives about one third of the critical density \cite{Turner}, requiring additional unknown component.

Alternatives to dark energy, discussed in the literature, include modification of gravity, for review see e.g. \cite{Mannheim}, and back-reaction of density perturbations on the Friedmann background \cite{Kolb,timescape}.



However, this standard theory does not take into account traces of the strong gravitational
waves of cosmological origin left in space. In the present paper we propose
the point of view that such traces can be a cause for the aforementioned
discrepancy (if so there is no need to search for any enigmatic substance
filling the universe). The sources of the long-lived cosmological waves are
the solitonic type inhomogeneities of the gravitational field near the Big
Bang. In general the inhomogeneities are unavoidable near the initial
cosmological singularity and some subset of them has the solitonic
structure. Due to expansion of the universe these inhomogeneities decay but
expel solitonic gravitational waves which also decay in course of
propagation through the expanding space transferring, however, their
energies to the Friedmann background deforming it and making the distances
different compared with those which would be observed without such waves 
\footnote{The phenomenon we describe here is due to the strong non-linear
gravitational waves of cosmological origin, however, it is akin to the
Zeldovich-Polnarev \cite{ZP} memory effect produced by the weak linear
gravitational waves generated by the colliding astrophysical objects. }.
The gravitational solitons are most important among all possible types of
waves because such disturbances are more stable and survive longer time than
others in the course of expansion. This effect has been described in the
paper \cite{B1} by example of single-soliton cylindrical wave propagating on
the Friedmann background where the background is supported by the matter in the
form of massless scalar field $\varphi $. The particular kind of
matter is of a little importance since constructed solutions contain only
gravitational waves and matter field remains unperturbed, it serves only to
support the background. In the case of flat model the background solution of
the Einstein equations in cylindrical coordinates is:%
\begin{equation}
-ds^{2}=t(-dt^{2}+dz^{2}+z^{2}dx^{2}+dy^{2})\text{ },\varphi =\left(
3/2\right) ^{1/2}\ln t\text{ },\text{ }  \label{1}
\end{equation}%
and exact solution \cite{B1} containing one gravitational soliton
propagating on this background has the form:%
\begin{align}
-ds^{2}= & f\left( t,z\right) \left( -dt^{2}+dz^{2}\right) \nonumber \\
+ & g_{11}\left(
t,z\right) dx^{2}+g_{22}\left( t,z\right) dy^{2}+2g_{12}\left( t,z\right)
dxdy\text{ },  \label{2}
\end{align}%
with the same unperturbed scalar field $\varphi $ as in (\ref{1}). Here $%
\left( x^{0},x^{1},x^{2},x^{3}\right) =\left( t,x,y,z\right) $ and metric
has the following structure:

\begin{align}
g_{11}= & \frac{t}{s^{2}t^{2}+\left( t^{2}+\mu \right) ^{2}} \nonumber \\ 
\times & \left[
s^{2}t^{2}z^{2}+z^{2}\left( t^{2}+\mu \right) ^{2}+qz^{2}\left( t^{2}+\mu
\right) -q^{2}\mu \right] \text{ },  \label{3}
\end{align}%
\begin{equation}
g_{22}=\frac{t}{s^{2}t^{2}+\left( t^{2}+\mu \right) ^{2}}\left[
s^{2}t^{2}+\left( t^{2}+\mu \right) ^{2}-q\left( t^{2}+\mu \right) \right] 
\text{ },  \label{4}
\end{equation}%
\begin{equation}
g_{12}=\frac{tqs\mu }{s^{2}t^{2}+\left( t^{2}+\mu \right) ^{2}}\text{ },
\label{5}
\end{equation}%
\begin{equation}
f=\frac{tl^{2}[s^{2}t^{2}+\left( t^{2}+\mu \right) ^{2}]}{%
s^{2}[l^{2}t^{2}+\left( t^{2}+\mu \right) ^{2}]}  \label{6}
\end{equation}%
In these formulas $l$ and $s$ are two arbitrary real constants and%
\begin{equation}
q=s^{2}-l^{2}\text{ }.  \label{7}
\end{equation}%
The function $\mu $ (representing the pole trajectory in the inverse
scattering method by which this solution was found) is:%
\begin{equation}
\mu =-\frac{1}{2}\left( l^{2}+t^{2}+z^{2}\right) +\frac{1}{2}\left[ \left(
l^{2}+t^{2}+z^{2}\right) ^{2}-4t^{2}z^{2}\right] ^{1/2}\text{ },  \label{8}
\end{equation}%
It is easy to see that everywhere in space-time the square root in this
expression is real and without loss of generality we can define it as
positive. In the limit when parameter $q$ tends to zero ($s^{2}\rightarrow
l^{2}$) the solution gives the background metric (\ref{1}). It is this fact that
permits to interpret the solution as an exact solitonic gravitational
perturbation on the Friedmann background. It is convenient to characterize
the solitonic field as exact deviation of the metric (\ref{2}) from its
background, which we designate by the upper index $(0)$. For the $%
g $-components of the metric (\ref{2}) this field is represented by the
symmetric matrix $H_{ab}$ ($a,b=1,2)$:%
\begin{align}
H_{11}= & \left( g_{11}-g_{11}^{\left( 0\right) }\right) \left( g_{11}^{\left(
0\right) }\right) ^{-1}, \\
H_{22}= & \left( g_{22}-g_{22}^{\left(
0\right) }\right) \left( g_{22}^{\left( 0\right) }\right) ^{-1},\label{8-A} \\
H_{12}= & H_{21}=g_{12}\left( g_{11}^{\left( 0\right) }g_{22}^{\left( 0\right)
}\right) ^{-1/2},  \label{8-B}
\end{align}%
and in the same way can be defined the perturbation $F$ for the $f$%
-component of the metric (\ref{2}):%
\begin{equation}
F=\left( f-f^{\left( 0\right) }\right) \left( f^{\left( 0\right) }\right)
^{-1}  \label{8-C}
\end{equation}%
Then in case of the background (\ref{1}) we have:%
\begin{align}
H_{11}= & \frac{g_{11}-tz^{2}}{tz^{2}}, \quad H_{22}=\frac{g_{22}-t}{t} \nonumber \\
H_{12}= & \frac{g_{12}}{tz}, \quad F=\frac{f-t}{t}.
\label{8-D}
\end{align}

\begin{figure}[ht!]
	\centering
		\includegraphics[width=\textwidth]{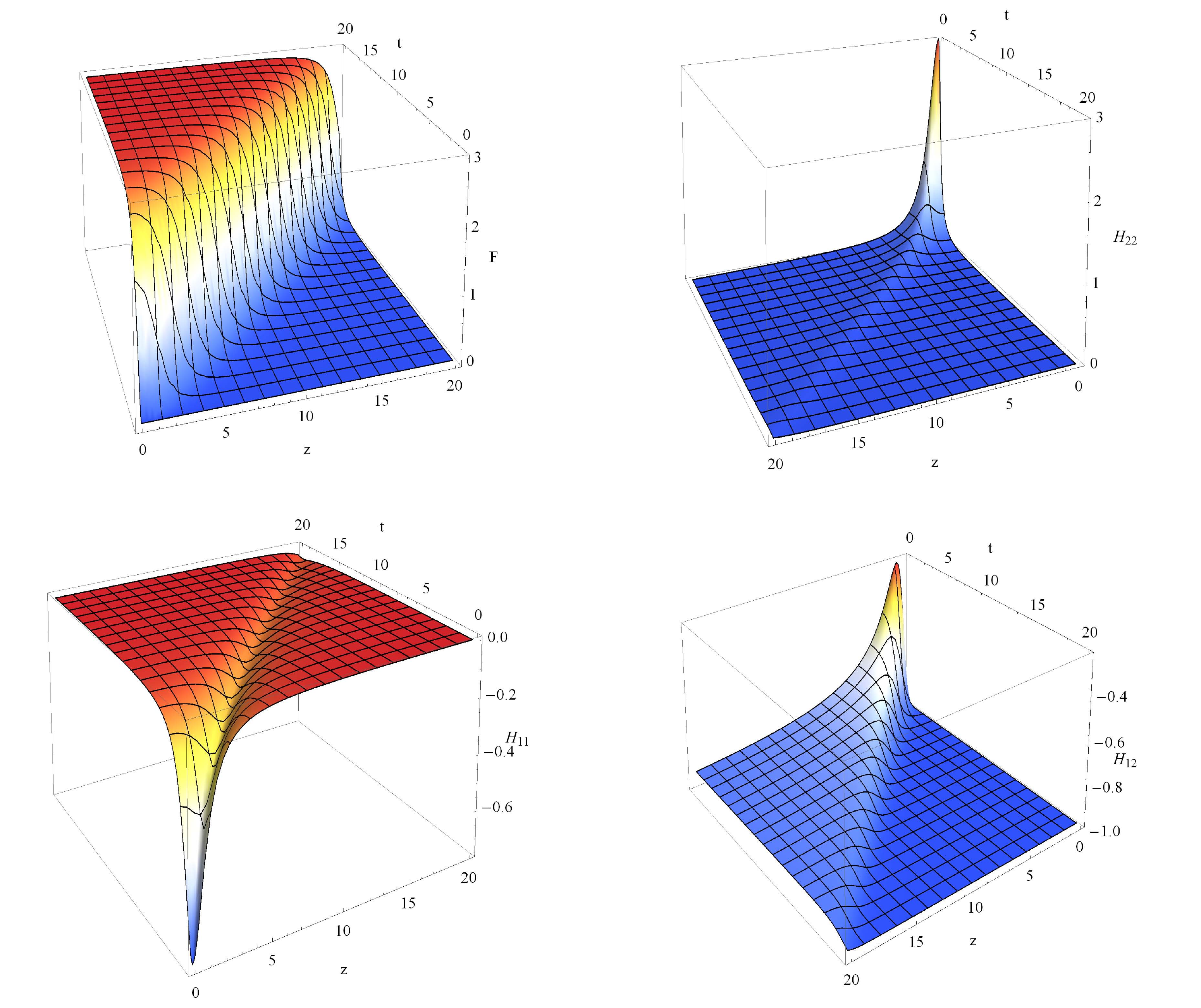}
	\caption{Single-soliton solution with parameters $l=1$, $s=1/2$ as a function of time $t$ and spatial coordinate $z$. From left to right and from top to bottom the following functions are shown: $F$, $H_{22}$, $H_{11}$ and $H_{12}$.}
	\label{case1}
\end{figure}

The solution (\ref{3})-(\ref{8}) contains no singularities other than the
usual cosmological singularity already present in the background (\ref{1}%
). The axis of cylindrical symmetry $z=0$ is regular. All fields and its
derivatives on the axis take some finite values. The solitonic excitations
at the initial time $t=0$ are concentrated around the axis $z=0$ with the
width $\delta z\sim l$ and disappear at infinity $z\rightarrow \infty $.
Hence, observers at the initial stage of expansion located far enough from
the axis see no difference from the usual Friedmann universe with metric (\ref%
{1}). During expansion the disturbance $H_{ab}$ around the axis vanish
but produces a solitonic gravitational wave moving away from the axis to
infinity, with amplitude decreasing in time. This wave propagates along
light-like line (in two-dimensional section) $t=z$ inside the strip with the
same width $\delta z\sim l$ as initial disturbances have (see Fig. \ref{case1}). 
It turns out that for the same observers located far from the axis but at the
final stages of the expansion, that is after the wave passed the region
around them, the solution describes again the flat Friedmann model, however,
with different measure for the time intervals and space distances in the
longitudinal $z$-direction (measures of the distances in the transversal
directions $x$ and $y$ do not change). From (\ref{8}) it follows that at $%
t\rightarrow 0$ the function $\mu $ tends to zero as $t^{2}$ and in this
limit for the metric coefficient $f$ \ (\ref{6}) we have $f=t$. In the limit 
$t\rightarrow \infty $ (and $t\gg z$) the function $\mu $ in the main
approximation does not depend on time and at the late phases of expansion we
have $f=$ $tl^{2}s^{-2}$. Consequently, passing of a cosmological
gravitational wave through the universe makes the scale factor in the
measure of the longitudinal distances for the initial and final Friedmann
backgrounds different because perturbation $F$ of this factor makes the jump
(see Fig. \ref{case1}): 
\begin{equation}
\lim_{t\rightarrow 0}F=0\text{ },\text{ }\lim_{t\rightarrow \infty ,\text{ }%
t\gg z}F=\frac{l^{2}-s^{2}}{s^{2}}\text{ }.  \label{8a}
\end{equation}%
This jump disappears if $l=s$ that is when the constant defined in $q$ (\ref{7}) vanish,
but in such case, as we already mentioned, the solution (\ref{2})-(\ref{6})
coincide identically with the background (\ref{1}), that is the case $l=s$
means the absence of any waves in the course of evolution.

Now the reasonable question to ask is whether such longitudinal memory
effect is only due the cylindrical symmetry and single-solitonic structure
of the chosen solution but can be absent in the more general cases. Let's
show that the same effect arises in the solutions containing
double-solitonic waves and no matter under which symmetry, cylindrical or
planar. These facts suggest a hint that the longitudinal memory effect is the
general phenomenon of solitonic waves propagating on an isotropic
homogeneous cosmological background. Some general consequences of this
phenomenon we will discuss in the section "Summary".

\section{Double-soliton gravitational waves on the Friedmann background}

\subsection{Cylindrical symmetry}

\begin{figure}[ht!]
	\centering
		\includegraphics[width=\textwidth]{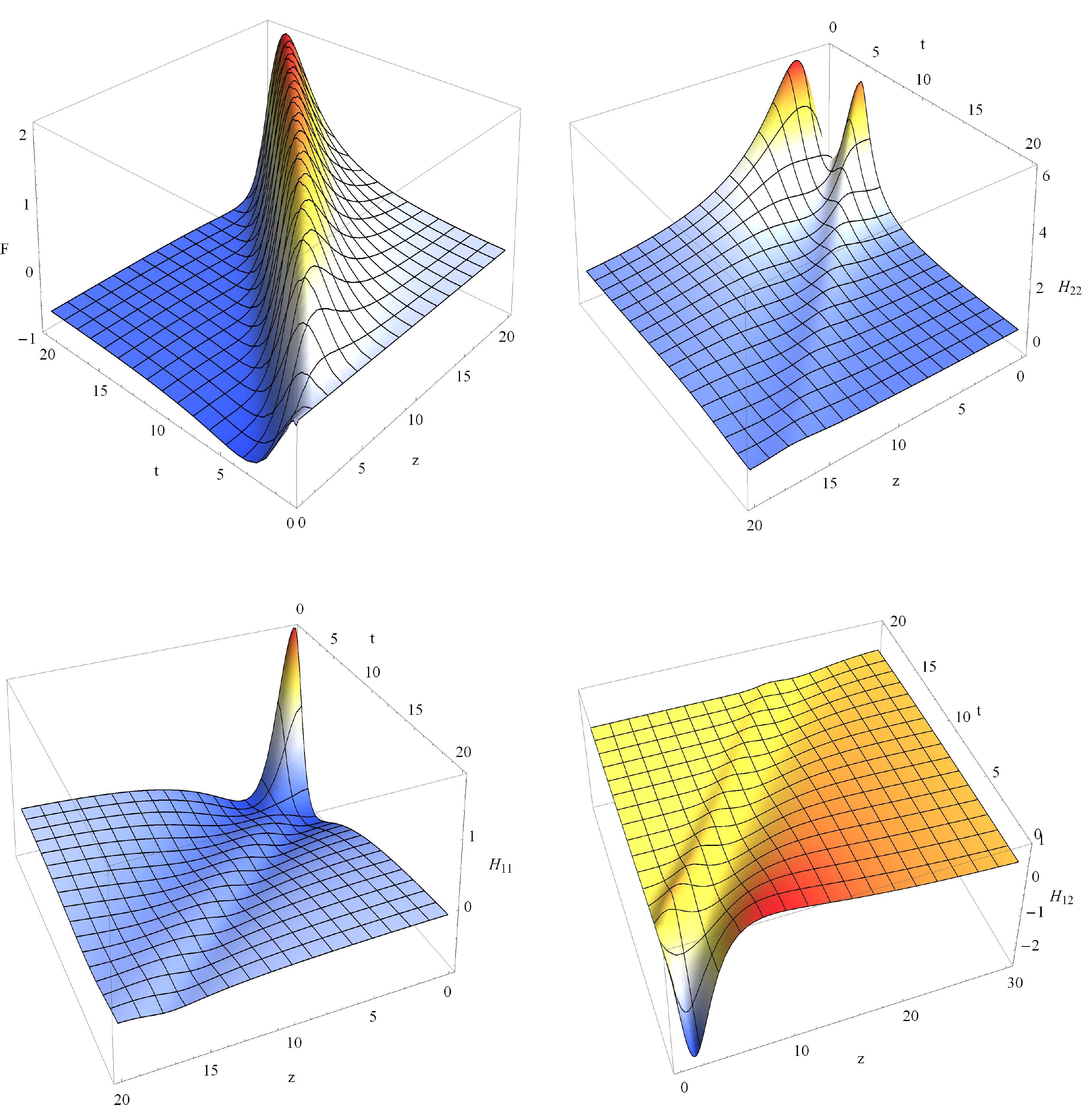}
	\caption{Double-soliton solution with parameters $l_1=-3$, $l_2=-2.8$, $\lambda_1=1.1$, $\lambda_2=1.2$ as a function of time $t$ and spatial coordinate $z$. From left to right and from top to bottom the following functions are shown: $F$, $H_{22}$, $H_{11}$ and $H_{12}$.}
	\label{case2}
\end{figure}

Using the same method we can add to Friedmann background (\ref{1}) two
solitonic gravitational waves of the same type as in the previous case. The
matter field $\varphi $ is again unperturbed. The corresponding exact
solution is described in Appendix I and here we outline only its properties
relevant to the effect we are interested in. Qualitatively these properties
are the same as in the previous single-soliton case. Again, the
inhomogeneous solitonic perturbations at the initial time $t=0$ are
concentrated around the axis of cylindrical symmetry and during expansion
decay but creates a gravitational waves moving along the light-like line $%
z=t $ away from the axis to infinity with amplitudes decreasing in time (see
Fig. \ref{case2}). 
For an region located far from the axis and at the initial stage of
expansion (that is before the waves reached this region) the metric is as in
Friedmann background (\ref{1}) but at the final stages of the expansion
(after the waves passed this region) the metric again take the Friedmann
form, however, with different measure for the time intervals and distances
in the longitudinal $z$-direction (measures of the distances in the
transversal directions $x$ and $y$ again do not change). Calculations show
that in the limit $t\rightarrow 0$ we have $f=t$ but when $t\rightarrow
\infty $ (and $t\gg z$) the scale factor $f$ becomes $f=t(l_{1}^{2}\lambda
_{1}-l_{2}^{2}\lambda _{2})^{2}\left[ l_{1}l_{2}\left( \lambda _{1}-\lambda
_{2}\right) \right] ^{-2}$, where $l_{1},l_{2},\lambda _{1},\lambda _{2}$
are four arbitrary constants the solution contains (see Appendix I). Again,
the perturbation $F$ of this factor makes the jump (see Fig. \ref{case2}):%
\begin{equation}
\lim_{t\rightarrow 0}F=0\text{ },\text{ }\lim_{t\rightarrow \infty ,\text{ }%
t\gg z}F=\frac{\left( l_{1}^{2}-l_{2}^{2}\right) (l_{1}^{2}\lambda
_{1}^{2}-l_{2}^{2}\lambda _{2}^{2})}{l_{1}^{2}l_{2}^{2}\left( \lambda
_{1}-\lambda _{2}\right) ^{2}}\text{ }.  \label{8b}
\end{equation}%
This jump vanish only in the limit $l_{1}^{2}\rightarrow l_{2}^{2}$ but, as
shown in Appendix I, in this limit any waves disappear and solution coincide
identically with the background Friedmann solution (\ref{1}). Then we have
the same phenomenon as in the previous single-soliton case.

\subsection{Plane wave symmetry}

The previous two cases describe cylindrical gravitational waves in the
Friedmann universe. To construct the plane waves on the Friedmann background we
need to represent it by Cartesian 3-dimensional space coordinates:%
\begin{equation}
-ds^{2}=t(-dt^{2}+dz^{2}+dx^{2}+dy^{2})\text{ },\text{ }\varphi =\left(
3/2\right) ^{1/2}\ln t\text{ }.  \label{8c}
\end{equation}%
However, for the real pole trajectories (real functions $\mu $ as in the
previous two cases) the inverse scattering method for this form of
background produce solutions with discontinuities of the first derivatives
of the metric tensor on the light cone $t^{2}=z^{2}$. This phenomenon
is related to the shock waves and need an appropriate physical interpretation.
Nevertheless, the wide set of fully analytical solutions also exists. This
set represents the gravitational solitons which correspond to an arbitrary
number of complex-conjugated pairs of the poles in the inverse scattering
machinery (that is an arbitrary number of complex-conjugated pairs of
functions $\mu $). In the present paper we consider the simplest case when
there is only one such pair $\mu _{1}=\rho e^{i\psi }$ $,$ $\mu _{2}=\rho
e^{-i\psi }$. The exact form of solution is given in Appendix II and omit
details of its derivation (for guide see \cite{BV}, subsection 2.2).

\begin{figure}[ht!]
	\centering
		\includegraphics[width=\textwidth]{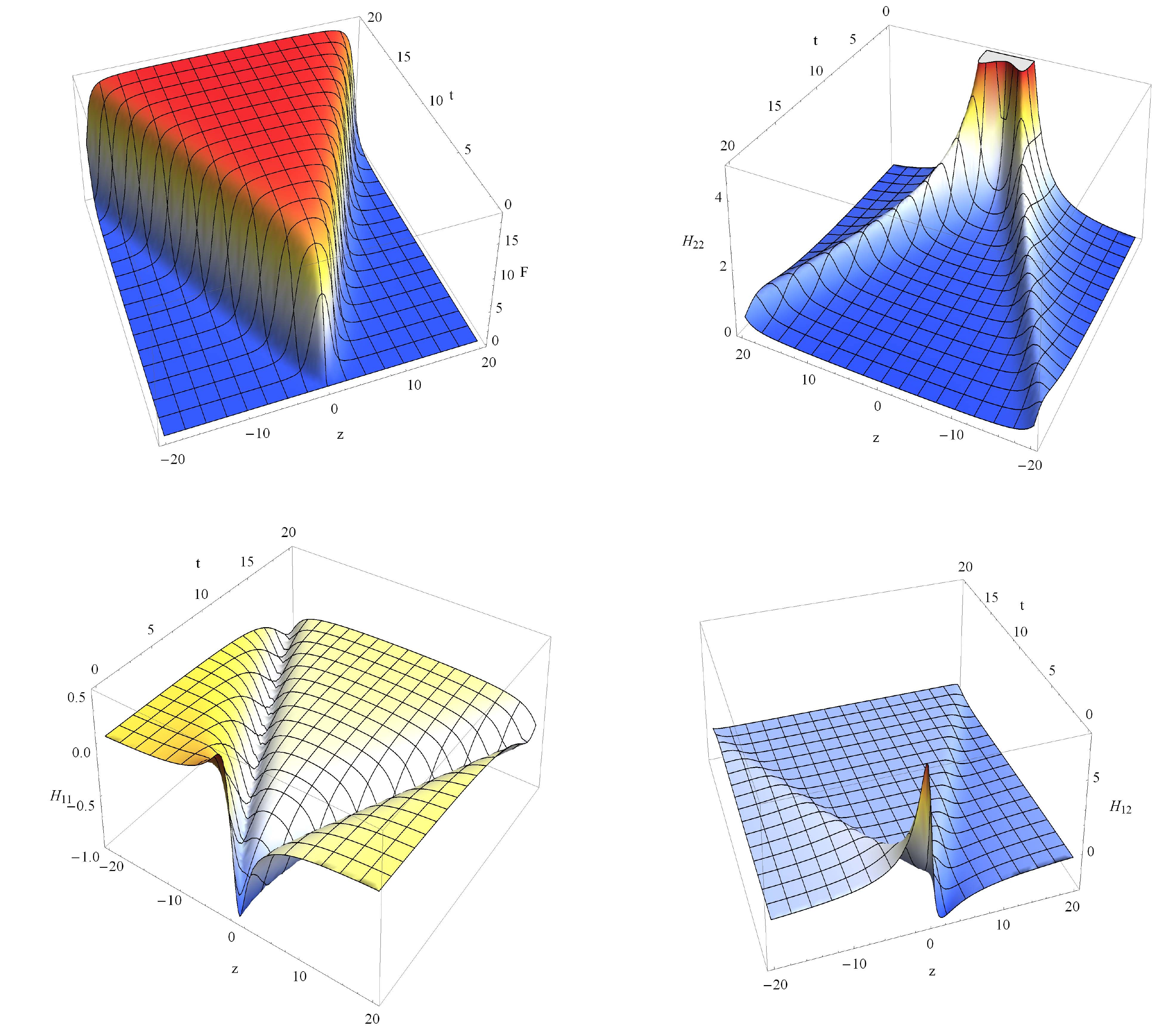}
	\caption{Double-soliton solution with plane wave symmetry having parameters $\omega_0=1$, $\sigma_0=2$, $\tau_0=-1$ as a function of time $t$ and spatial coordinate $z$. From left to right and from top to bottom the following functions are shown: $F$, $H_{22}$, $H_{11}$ and $H_{12}$. Function $H_{22}$ is finite everywhere, but its peak is not shown for better illustration of wave propagation.}
	\label{case3}
\end{figure}

For the plane wave case the solitonic fields we define by the same formulas (%
\ref{8-A})-(\ref{8-C}) but these perturbations take a different
functional form with respect to the previous two cases since the background (%
\ref{8c}) is different:%
\begin{equation}
H_{11}=\frac{g_{11}-t}{t}\text{ },\text{ }H_{22}=\frac{g_{22}-t}{t},\text{ }%
H_{12}=\frac{g_{12}}{t}\text{ },\text{ }F=\frac{f-t}{t}  \label{8d}
\end{equation}%
From qualitative point of view the regular plane solitonic waves behave
similarly as in the preceding cases for the cylindrical solutions. The
inhomogeneous solitonic perturbation at the initial time $t=0$ is
concentrated around the central plane $z=0$ of the 3-dimensional space and
during expansion decays but creates two gravitational waves moving (one to
the left and other to the right inside some strips along the light cone $%
z^{2}=t^{2}$) away from the central plane to infinities $z\rightarrow \pm
\infty $ with amplitudes decreasing in time, see Fig. \ref{case3}. At the initial stage of
expansion for the left and right regions located far from the central
disturbance (that is before the waves reached these regions) the metric is
as in Friedmann background (\ref{8c}) but at the final stages of the
expansion (after the waves passed these regions) the metric again takes the
Friedmann form, however, with different measure for the time intervals and
distances in the longitudinal $z$-direction (measures of the distances in
the transversal directions $x$ and $y$ again do not change). Calculations
show that in both left and right far regions the metric coefficient $f$ near
Big Bang $t\rightarrow 0$ is the same as in background, that is $f=t$, but
at the late stages of expansion $t\rightarrow \infty $ (and $t\gg z$) it
takes the value $f=t\cosh ^{2}\sigma _{0}\left( \sin ^{2}\tau _{0}\right)
^{-1}$, where $\sigma _{0}$ and $\tau _{0}$ are arbitrary real constants the
solution depends on (see Appendix II). Again, the perturbation $F$ of the
metric coefficient $f$ makes the jump (see Fig. \ref{case3}):%
\begin{equation}
\lim_{t\rightarrow 0}F=0\text{ },\text{ }\lim_{t\rightarrow \infty ,\text{ }%
t\gg z}F=\frac{\cosh ^{2}\sigma _{0}-\sin ^{2}\tau _{0}}{\sin ^{2}\tau _{0}}%
\text{ },  \label{8e}
\end{equation}%
This jump vanishes only in that special case when $\sigma _{0}=0$ and $\tau
_{0}=\pi /2$ but, as follows from the exact form of the solution (see
Appendix II), in this case the waves disappear and solution gives
identically the background Friedmann metric (\ref{8c}).

What is new and intriguing here (with respect to the two preceding cases)
is the fact that for the plane waves (corresponding to the
complex-conjugated pair of the pole trajectories in the inverse scattering
method) the memory effect always increases the time intervals and
longitudinal distances (because $\cosh ^{2}\sigma _{0}\geqslant \sin
^{2}\tau _{0}$ for any values of the arbitrary constants).

\section{Summary}

The crucial point in each complicated physical process is to find the basic
elementary phenomenon. We hope that by above analysis we identified such
basic point in the problem of excess of the value of the scale factor over
its usual Friedmann prescription. One may expect in the real universe a
huge amount chaotically distributed inhomogeneous solitonic perturbations
near the Big Bang and in the course of the expansion all of them will decay
sending the gravitational waves to the space, each in different direction.
Solitons do not prevent each other to propagate keeping their shapes and
directions of propagations. All of them will change their longitudinal
distances, however, some direction in space can be transversal for one wave
but longitudinal for the other. As a result the distances will be changed
equally in any direction (probably with some anisotropy). The examples with
real functions $\mu $ (real poles in the inverse scattering technique) show
that distances during expansion can be either increasing or decreasing which
depends on the values of the arbitrary constants in the expressions (\ref{8a}%
) and (\ref{8b}). But complex $\mu $ (complex poles) produce effect only
with increasing distances: the ratio of the metric coefficients $f$ after
the waves passed to its value before the waves came is always greater than
unity. The last property probably is more preferable from the point of view
of observations and it looks interesting that it can be given some support
of statistical flavour. The main step in construction of solitonic solution
is to choose the position of poles (that is the character of the functions $%
\mu $, real or complex) of the dressing matrix in the complex plane of the
spectral parameter. Of course, we cannot have any information on how many
initial cosmological disturbances can correspond to the complex poles and
how many to real. In the absence of such information we are forced to
evaluate the relative amounts of the number of poles just by the volumes of
those parts of the complex plane of the spectral parameter where
corresponding poles can be located. It is evident that the number of the
complex poles should dominate essentially because a line (real axis of the
complex plane) represents a set of measure zero with respect to the whole
plane. Then in the mixture of the solitonic cosmological waves the dominant contribution will come from those, which increase the distances and the global average effect also will correspond to the increasing of the time interval and
space distances. This new effect may provide alternative explanation for apparent cosmic acceleration inferred from cosmological observations.

\onecolumngrid

\section{Appendix I}

Consider the case when we have two real pole trajectories in the inverse
scattering machinery and correspondingly two functions $\mu _{1}\left(
t,z\right) $ and $\mu _{2}\left( t,z\right) $ : 
\begin{equation}
\mu _{1}=-\frac{1}{2}\left( l_{1}^{2}+t^{2}+z^{2}\right) +\frac{1}{2}\left[
\left( l_{1}^{2}+t^{2}+z^{2}\right) ^{2}-4t^{2}z^{2}\right] ^{1/2}\text{ },
\label{9}
\end{equation}%
\begin{equation}
\mu _{2}=-\frac{1}{2}\left( l_{2}^{2}+t^{2}+z^{2}\right) +\frac{1}{2}\left[
\left( l_{2}^{2}+t^{2}+z^{2}\right) ^{2}-4t^{2}z^{2}\right] ^{1/2}\text{ },
\label{10}
\end{equation}%
where $l_{1\text{ }}$and $l_{2}$ are two arbitrary real constants.
Everywhere in space-time the square roots in these expressions are real and
without loss of generality we can define them as positive (taking the
different signs in front of these roots does not change the basic results we
are interested in). Apart from $l_{1\text{ }}$and $l_{2}$ another two
arbitrary real constants appear in the solution which constants we denote as 
$\lambda _{1}$ and $\lambda _{2}$. Now the interval has the same form as in (%
\ref{2}) but with more cumbersome metric coefficients: 
\begin{equation}
g_{11}=U+\gamma ^{2}V+2\gamma W\text{ },\text{ }g_{22}=V\text{ },\text{ }%
g_{12}=W+\gamma V\text{ },  \label{11}
\end{equation}%
where by $\gamma $ we denote the constant%
\begin{equation}
\gamma =-\frac{l_{1}^{2}-l_{2}^{2}}{l_{1}^{2}\lambda _{1}-l_{2}^{2}\lambda
_{2}}\text{ },  \label{12}
\end{equation}%
and quantities $U,V,W$ are\footnote{%
It is worth mentioning that $g_{11}=U$ $,$ $g_{22}=V$ $,$ $g_{12}=W$ with
the same $f$ and $\varphi $ from (\ref{18})-(\ref{18a}) also is a solution
of the Einstein equations but the regular behavior of the solution on the
axis $z=0$ demands transformation of the dummy coordinates $x,y$ as $%
x\rightarrow x,$ $y\rightarrow y+\gamma x$. This produces the metric
coefficients in the form (\ref{11}) as linear combinations of quantities $%
U,V,W$.}:%
\begin{equation}
U=\frac{tz^{2}}{2Q}\left[ 2Q+M_{1}^{2}+M_{2}^{2}+M_{1}^{2}M_{2}^{2}+1+\frac{%
\mu _{1}+\mu _{2}}{\mu _{1}-\mu _{2}}\left( M_{1}^{2}-M_{2}^{2}\right) -%
\frac{t^{2}z^{2}+\mu _{1}\mu _{2}}{t^{2}z^{2}-\mu _{1}\mu _{2}}%
(M_{1}^{2}M_{2}^{2}-1)\right] \text{ },  \label{13}
\end{equation}%
\begin{equation}
V=\frac{t}{2Q}\left[ 2Q+M_{1}^{2}+M_{2}^{2}+M_{1}^{2}M_{2}^{2}+1-\frac{\mu
_{1}+\mu _{2}}{\mu _{1}-\mu _{2}}\left( M_{1}^{2}-M_{2}^{2}\right) +\frac{%
t^{2}z^{2}+\mu _{1}\mu _{2}}{t^{2}z^{2}-\mu _{1}\mu _{2}}\left(
M_{1}^{2}M_{2}^{2}-1\right) \right] \text{ },  \label{14}
\end{equation}%
\begin{equation}
W=\frac{tz}{2Q}\left[ \frac{\mu _{1}+\mu _{2}}{\mu _{1}-\mu _{2}}\left(
M_{1}-M_{2}\right) \left( M_{1}M_{2}-1\right) +\frac{t^{2}z^{2}+\mu _{1}\mu
_{2}}{t^{2}z^{2}-\mu _{1}\mu _{2}}\left( M_{1}+M_{2}\right) \left(
M_{1}M_{2}+1\right) \right] \text{ },  \label{15}
\end{equation}%
\begin{equation}
Q=\mu _{1}\mu _{2}\left[ \frac{t^{2}z^{2}}{(t^{2}z^{2}-\mu _{1}\mu _{2})^{2}}%
\left( M_{1}M_{2}+1\right) ^{2}+\frac{1}{\left( \mu _{1}-\mu _{2}\right) ^{2}%
}\left( M_{1}-M_{2}\right) ^{2}\right] \text{ },  \label{16}
\end{equation}%
\begin{equation}
M_{1}=\frac{\left( z^{2}+\mu _{1}\right) \lambda _{1}}{z}\text{ },\text{ }%
M_{2}=\frac{\left( z^{2}+\mu _{2}\right) \lambda _{2}}{z}\text{ }.
\label{17}
\end{equation}%
The metric coefficient $f$ is:%
\begin{equation}
f=\frac{t(l_{1}^{2}-l_{2}^{2})^{2}\lambda _{1}\lambda _{2}\mu _{1}\mu _{2}Q}{%
(\lambda _{1}-\lambda _{2})^{2}(t^{2}z^{2}-\mu _{1}^{2})(t^{2}z^{2}-\mu
_{2}^{2})M_{1}M_{2}}\text{ },  \label{18}
\end{equation}%
and scalar matter field $\varphi $ remains the same as in background:%
\begin{equation}
\varphi =\left( 3/2\right) ^{1/2}\ln t\text{ }.  \label{18a}
\end{equation}

The complexity of this exact solution creates no serious difficulties
because we need to know only few basic properties of the solution and its
asymptotic behavior for large and small values of coordinates. These
properties and asymptotics can be established easily. First of all in the
limit $l_{1}^{2}\rightarrow l_{2}^{2}$ solution again turns to the Friedmann
background (\ref{1}). [To see this it is necessary to expand all terms with
respect to the difference $l_{1}^{2}-l_{2}^{2}$ assuming it infinitesimally
small, in such limit $\mu _{1}-\mu _{2}\sim $ $l_{1}^{2}-l_{2}^{2}$ and $%
Q\sim \left( l_{1}^{2}-l_{2}^{2}\right) ^{-2}$]. This means that solution
indeed can be interpreted as representing the exact gravitational waves
propagating on the flat Friedmann background.

All other properties of the solution (\ref{9})-(\ref{18a}) are very similar
to those for the single-soliton case. The solution contains no singularities
other than the usual cosmological singularity already presented in the
background. The axis of cylindrical symmetry $z=0$ is regular. All fields
and its derivatives on the axis take some finite values. The gravitational
perturbations at initial time $t=0$ are concentrated around the axis $z=0$
with the widths in $z$-direction of order of unity if constants $l_{1},$ $%
l_{2},$ $\lambda _{1},$ $\lambda _{2}$ are chosen to be of the order of
unity. Other relevant properties of this solution we described already in
the text.

\subsection{Appendix II}

To construct the solitonic plane waves on the Friedmann background (\ref{8c}) we
apply the standard inverse scattering procedure corresponding to the pair of
complex conjugated poles ( $\mu _{1}=\rho e^{i\psi }$ $,$ $\mu _{2}=\rho
e^{-i\psi }$) which results in the solution of the following form:

\begin{equation}
g_{11}=\frac{t}{D}\left[ D+\cos 2\tau _{0}+\cosh 2\sigma _{0}+\frac{z\left(
t^{2}-\rho ^{2}\right) }{w_{0}\left( t^{2}+\rho ^{2}\right) }\sin 2\tau _{0}-%
\frac{t^{2}+\rho ^{2}}{t^{2}-\rho ^{2}}\sinh 2\sigma _{0}\right] \text{ },
\label{21}
\end{equation}%
\begin{equation}
g_{22}=\frac{t}{D}\left[ D+\cos 2\tau _{0}+\cosh 2\sigma _{0}-\frac{z\left(
t^{2}-\rho ^{2}\right) }{w_{0}\left( t^{2}+\rho ^{2}\right) }\sin 2\tau _{0}+%
\frac{t^{2}+\rho ^{2}}{t^{2}-\rho ^{2}}\sinh 2\sigma _{0}\right] \text{ },
\label{22}
\end{equation}%
\begin{equation}
g_{12}=\frac{2t}{D}\left[ \frac{z\left( t^{2}-\rho ^{2}\right) }{w_{0}\left(
t^{2}+\rho ^{2}\right) }\sinh \sigma _{0}\sin \tau _{0}+\frac{t^{2}+\rho ^{2}%
}{t^{2}-\rho ^{2}}\cosh \sigma _{0}\cos \tau _{0}\right] \text{ },
\label{23}
\end{equation}%
\begin{equation}
D=\frac{(t^{2}-\rho ^{2})^{4}\sin ^{2}\tau _{0}+16w_{0}^{2}t^{2}\rho
^{4}\cosh ^{2}\sigma _{0}}{4w_{0}^{2}\rho ^{2}(t^{2}-\rho ^{2})^{2}}\text{ },
\label{24}
\end{equation}%
\begin{equation}
f=\frac{t\left[ (t^{2}-\rho ^{2})^{4}\sin ^{2}\tau _{0}+16w_{0}^{2}t^{2}\rho
^{4}\cosh ^{2}\sigma _{0}\right] }{\left[ (t^{2}-\rho
^{2})^{4}+16w_{0}^{2}t^{2}\rho ^{4}\right] \sin ^{2}\tau _{0}}\text{ },
\label{25}
\end{equation}%
where $\sigma _{0},$ $\tau _{0}$ and $w_{0}$ are three real arbitrary
constants. The function $\rho \left( t,z\right) $ is real and positive and
follows from the equation:%
\begin{equation}
\frac{4z^{2}\rho ^{2}}{\left( t^{2}+\rho ^{2}\right) ^{2}}+\frac{%
4w_{0}^{2}\rho ^{2}}{\left( t^{2}-\rho ^{2}\right) ^{2}}=1\text{ }.
\label{26}
\end{equation}%
The matter field, as in the previous cases, is $\varphi =\left( 3/2\right)
^{1/2}\ln t$ that is remains unperturbed. The algebraic equation (\ref{26})
has two real and positive solutions for function $\rho \left( t,z\right) $.
We chosed that one which in the limit $t\rightarrow 0$ has asymptotcs $\rho
=t^{2}\left[ 4\left( w_{0}^{2}+z^{2}\right) \right] ^{-1/2}$ and in the
limit $t\rightarrow \infty $ tends to $\rho =t-w_{0}$ (second root for $\rho 
$ gives the same physical results).

It is easy to see that there are two different ways in the parameter space
to get the background limit. First is to take $\sigma _{0}=0$ and $\tau
_{0}=\pi /2$ in which case solution gives the seed metric (\ref{8c}). The
second and independent possibility is to take limit $w_{0}\rightarrow 0$
under which solution also tends to coincide with this background. Then
solution indeed represents the two exact solitonic plane gravitational wave
propagating on the Friedmann universe.

\onecolumngrid
\twocolumngrid

\end{document}